\documentclass[12pt,preprint]{aastex}

\begin{document}
\title{Recent Star Formation in the Extreme Outer Disk of M83}
\author{
David A. Thilker\altaffilmark{1},
Luciana Bianchi\altaffilmark{1}, 
Samuel Boissier\altaffilmark{2},
Armando Gil de Paz\altaffilmark{2},
Barry F. Madore\altaffilmark{2},
D. Christopher Martin\altaffilmark{3},
Gerhardt R. Meurer\altaffilmark{1},
Susan G. Neff\altaffilmark{4},
R. Michael Rich\altaffilmark{5},
David Schiminovich\altaffilmark{3},
Mark Seibert\altaffilmark{3},
Ted K. Wyder\altaffilmark{3},
Tom A. Barlow\altaffilmark{3},
Yong-Ik Byun\altaffilmark{6},
Jose Donas\altaffilmark{7},
Karl Forster\altaffilmark{3},
Peter G. Friedman\altaffilmark{3},
Timothy M. Heckman\altaffilmark{8},
Patrick N. Jelinsky\altaffilmark{9},
Young-Wook Lee\altaffilmark{6},
Roger F. Malina\altaffilmark{7},
Bruno Milliard\altaffilmark{7},
Patrick Morrissey\altaffilmark{3},
Oswald H. W. Siegmund\altaffilmark{9},
Todd Small\altaffilmark{3},
Alex S. Szalay\altaffilmark{8}, and
Barry Y. Welsh\altaffilmark{9}
}

\altaffiltext{1}{Center for Astrophysical Sciences, The Johns Hopkins
University, 3400 N. Charles St., Baltimore, MD 21218, dthilker@pha.jhu.edu}

\altaffiltext{2}{Observatories of the Carnegie Institution of Washington,
813 Santa Barbara St., Pasadena, CA 91101}

\altaffiltext{3}{California Institute of Technology, MC 405-47, 1200 East
California Boulevard, Pasadena, CA 91125}

\altaffiltext{4}{Laboratory for Astronomy and Solar Physics, NASA Goddard
Space Flight Center, Greenbelt, MD 20771}

\altaffiltext{5}{Department of Physics and Astronomy, University of
California, Los Angeles, CA 90095}

\altaffiltext{6}{Center for Space Astrophysics, Yonsei University, Seoul
120-749, Korea}

\altaffiltext{7}{Laboratoire d'Astrophysique de Marseille, BP 8, Traverse
du Siphon, 13376 Marseille Cedex 12, France}

\altaffiltext{8}{Department of Physics and Astronomy, The Johns Hopkins
University, Homewood Campus, Baltimore, MD 21218}

\altaffiltext{9}{Space Sciences Laboratory, University of California at
Berkeley, 601 Campbell Hall, Berkeley, CA 94720}



\begin{abstract}

Ultraviolet imaging with the {\it Galaxy Evolution Explorer} (GALEX)
has revealed an extensive sample of UV-bright stellar complexes in the
extreme outer disk of M83, extending to about four times the
radius where the majority of \ion{H}{2} regions are detected ($R_{HII} =
5.1\arcmin$ or 6.6 kpc).  These sources are typically associated with
large-scale filamentary \ion{H}{1} structures in the warped outer disk of M83,
and are distributed beyond the galactocentric radii at which molecular
ISM has yet been detected.  We present measured properties of these
stellar complexes, including FUV and NUV magnitudes and local gas
surface density.  Only a subset of the outer disk UV sources have
corresponding \ion{H}{2} regions detected in H$\alpha$ imaging,
consistent with a sample of mixed age in which some sources are a few
Myr old and others are much more evolved ($\sim 10^8$ yr). 

\end{abstract}




\section{Introduction}
\label{sintro}

The {\it Galaxy Evolution Explorer} (GALEX), with its $1.25\arcdeg$
field-of-view and sensitivity to stellar populations younger than a
few hundred Myr, is remarkably well-suited for addressing the topic of
star formation at large galactocentric radii. Recent star formation
within such environments has been detected in deep H$\alpha$ and
broadband observations of a few galaxies (NGC 628, NGC1058, NGC6946:
Ferguson et al. 1998a, also Lelievre \& Roy 2000 for NGC628; M31:
Cuillandre et al. 2001; NGC6822: de Blok \& Walter 2003).  One of the
goals of the GALEX Nearby Galaxy Survey (NGS: Bianchi et al. 2004) is
to conduct a census of outer disk star formation in a statistically
significant sample of galaxies.

The confirmed occurrence of star formation in the outer disk of
ordinary spiral galaxies has far-reaching implications.  First, if
massive stars are forming at large radii, then the outer disk
continues to undergo chemical enrichment.  This may clarify the origin of dust detected in the extended HI disks of spiral
galaxies (Popescu \& Tuffs 2003, A\&A 410L, 21P). Secondly, radiative
and mechanical feedback processes associated with massive stars are
expected to be more efficient in the
comparatively rarified interstellar medium (ISM) and uncrowded
environment at large radii.  In particular, the unresolved issue of
\ion{H}{1} production via H$_2$ dissociation (Smith et al. 2000, Allen
1986) can be studied in detail.

The presence of recently formed stellar complexes at large galactocentric
radii also provides a simplified laboratory for investigating the star
formation threshold, the (minimum) gas column density required for
star formation to occur spontaneously.  The precise workings of this
mechanism remain a matter of debate.  Kennicutt (1989) demonstrated
that star formation is rare in disk environments having total gas
surface density $\Sigma_{gas} < 5-10$ M$_\sun$ pc$^{-2}$.  Martin \&
Kennicutt (2001) showed the observed threshold gas density varies by
about an order of magnitude at the ``edge'' of the star-forming disk
($R = R_{HII}$) in different galaxies, but the ratio of $\Sigma_{gas}$
to a critical density (which depends on the physical and dynamical
conditions of each galaxy in particular) is approximately constant at
this position. 

Deep H$\alpha$ imaging by Ferguson et al. (1998a) and Lelievre \& Roy
(2000) showed that star-formation regions only a few Myr old exist
beyond two optical radii (R$_{25}$) in NGG~628, NGC~1058, and
NGC~6946.  Lelievre \& Roy (2000) demonstrated that the \ion{H}{2}
region luminosity function (LF) slope is significantly steeper in the
outer disk, relative to a galaxy-wide sample, possibly indicating a
genuine change in the initial cluster mass function (CMF) or a
population which is preferentially older.  Ferguson et al. (1998a)
noted that the outer \ion{H}{2} regions appear small, isolated, and
faint compared to inner disk analogs.

In this Letter, we present far-UV (FUV) and near-UV (NUV) wide-field
GALEX imagery which reveals an extensive population of recent
star-forming regions in the extreme outer disk of M83, out to $\sim
3\times$ the Holmberg radius.  Some of these sources were discovered
by Wamsteker et al. (1983) in deep red (127-04+RG630) plates, but
observed properties were never published.  We analyze the GALEX data
in conjuction with visible and $\lambda$21cm imagery, to place the outer disk
sources in context relative to the underlying stellar population and
extended \ion{H}{1} disk.  {\it The GALEX data is of particular importance
because it demonstrates that H$\alpha$ observations will fail to
detect a significant population of moderate age clusters in the
outermost disk of spiral galaxies, leading to an underestimate of
galactic evolution in these locales.}  Processes such as chemical
enrichment (Ferguson et al. 1998b), disk heating (Sellwood \& Balbus
1999), and ISM phase balance (Allen 1986) may be significantly
influenced, given that the (predominantly) B star population traced by
GALEX effectively drives these feedback mechanisms.  GALEX NGS
observations show that M83 is not unique in terms of outer disk UV
morphology.  We will analyze UV-visible-IR outer disk properties of a
representative galaxy sample in a subsequent paper.

\section{Observations and Data Analysis}

M83 (NGC5236) was observed with GALEX on 7 June 2003 for a single
orbital night as part of the NGS, with an exposure of 1352 seconds in
both FUV (1350--1750~\AA) and NUV (1750--2750~\AA).  Figure 1 shows
the color-composite GALEX image of M83.  Two bright stars eastward of
M83 prevented us from centering the target in the field, but our
data includes most of the outer disk.  RMS sensitivity is 26.6(26.8)
AB mag for FUV(NUV), or 27.5(27.6) AB mag arcsec$^{-2}$ expressed in
units of surface brightness evaluated at the scale of the PSF.

Boissier et al. (2004, this volume) also analyze the GALEX M83
dataset, emphasizing the inner disk in regard to extinction
and chemical evolution models.

Courtesy of P. Crosthwaite, we gained access to the $\lambda21$cm VLA
observations discussed by Tilanus \& Allen (1993).  For the inner
$14\arcmin\times14\arcmin$ of M83, we also used the CO map of
Crosthwaite et al. (2002) to estimate $\Sigma_{gas}$ as a function of
position.  Figures 1, 4, and 5 pertain to the \ion{H}{1} and CO
observations.  Narrowband H$\alpha$ and R-band imaging of M83 was
obtained using the Swope 1-m telescope at Las Campanas Observatory
(see Fig. 5)

We adopt a distance of 4.5 Mpc to M83 (Thim et al. 2003), thus the
$\sim 5\arcsec$ GALEX PSF corresponds to a linear scale of 110
pc.  Accordingly, GALEX detections in M83 are generally not single
clusters, but rather aggregates possibly consisting of a few discrete
(unresolved) clusters.

Photometry of the discrete UV sources within M83 was accomplished
using Source Extractor (Bertin \& Arnouts 1996) to define an isophotal
aperture for every object detected within our NUV image. These
irregular region boundaries were then applied to both the NUV and FUV
data for the measurement of background-subtracted source flux, after
applying a distortion correction to the FUV image to ensure
registration with our NUV data.  The background for every source was
estimated locally (within $\sim 250$ pc). We measured median surface
brightness as a function of galactocentric radius as described by
Bianchi et al. (2004, this volume) with $i = 25\arcdeg$ and PA =
$226\arcdeg$ (Crosthwaite et al. 2002).

\section{Recent star formation in the\\ outer disk of M83}

M83 is considered prototypical of those galaxies showing a
well-defined edge of the star-forming disk (eg. Martin \& Kennicutt
2001).  The surface density of \ion{H}{2} regions declines abruptly
near R$_{HII} = 5.1\arcmin$ (6.6 kpc) radius (cf. R$_{Holm} =
7.3\arcmin$, 9.6 kpc), suggesting that massive star formation is now
confined to M83's inner disk, despite the prominence of a rather
extended \ion{H}{1} disk (Rogstad et al. 1974, Huchtmeier \& Bohnenstengel
1981, see also Fig. 1).  Our GALEX observations modify this view.
While the distribution of UV-bright stellar complexes is highly
concentrated in the area occupied by \ion{H}{2} regions (tracing O
stars), many UV sources (tracing O and B stars) are easily
recognizable far beyond the ``edge'' of this distribution.
Furthermore, diffuse UV emission is also detected beyond the
optically-bright star-forming disk.

\subsection{Outer disk, UV-bright stellar complexes}

Our GALEX observations of M83 revealed $>100$ UV-bright
star-forming regions and stellar clusters beyond R$_{HII}$.  Figure 1
shows that the outer disk star-forming regions appear preferentially
located on local maxima or filaments in the structure of the warped
\ion{H}{1} disk.  The distribution of GALEX sources extends to the
limit of the field surveyed at the VLA by Tilanus \& Allen (1993).  We
suspect that additional M83 star-forming regions may be located in the
remainder of the \ion{H}{1} envelope (Huchtmeier \& Bohnenstengel
1981), but this is challenging to confirm without corroborating
\ion{H}{1} morphology.

Upon inspection of POSS2 images, we found that the majority of
the outer disk GALEX detections could be discerned in IIIaJ+GG385
(blue) plates.  However, the apparent contrast of the
sources is lower in the visible than at UV wavelengths.  An
undetermined fraction of the outer disk star-forming complexes were
discovered by Wamsteker et al. (1983) in their analysis of deep red
(127-04+RG630) Schmidt plates. To our knowledge, this is the only
mention of the outer disk sources in this galaxy prior to the GALEX data.

Figure 2 presents the FUV luminosity function for M83 star-forming
complexes in various disk locations.  The outer disk complexes (dashed
line) appear systematically fainter than their inner disk counterparts
(dotted line). The median FUV luminosity of the inner disk population
is a factor of 3 greater than the corresponding figure beyond
R$_{HII}$ (only correcting for foreground extinction).  Internal
extinction is likely more severe in the inner disk.  Thus, the
intrinsic luminosity difference may be even higher than that shown in
Fig 2, although we caution that blending and photometric
incompleteness (more significant at $R < R_{HII}$) would have an
opposite effect.  The shape of the apparent FUV LF differs between the
inner and outer disk populations. Our M83 FUV luminosity function of
outer disk stellar complexes is steeper than for the inner disk
population.  A similar trend was noted in NGC~628 for the H$\alpha$ LF
of \ion{H}{2} regions (Levievre \& Roy 2000).  Because our GALEX measurements
represent the aggregate flux of more than one independent cluster (or
simply an extended distribution of massive stars), it is premature to
interpret the precise slope of these FUV LFs in terms of the CMF. We
note that blending tends to flatten observed LFs, but the magnitude of
this effect may vary with environment.  High-resolution follow-up
observations would resolve ambiguities related to the FUV LF and
median luminosity estimates. Nevertheless, the maximum UV luminosity
exhibited by the population of outer disk sources implies an upper
limit of about $10^5 M_\sun$ for the most recently formed complexes
(see also Fig. 3).

We compare the observed (NUV, FUV--NUV) color magnitude diagram (CMD)
for UV-bright complexes with population synthesis models of Bruzual \&
Charlot (2003) in Figure 3.  M83's outer disk sources have FUV--NUV
colors generally consistent with ages from a few Myr up to 400 Myr and
are generally less massive than sources with $R < R_{HII}$.

Only a subset of the outer disk UV-bright sources have detected
counterparts in H$\alpha$ imagery.  This effect could be partially due
to a combination of limited sensitivity and the lower expected
emission measure of \ion{H}{2} nebulae in a low-density environment, but
supports the view that our GALEX observations trace a diverse
population, comprised of both young and moderate-age complexes.  A distinct advantage of UV observations is to directly trace stellar, rather than nebular, emission.

Figure 4 presents the distribution of total gas surface density at the
locations of all inner(outer) disk sources using a dotted(dashed)
line.  The local surface densities at which outer disk star-forming
complexes appear to be forming are typically subcritical, with respect
to the disk instability criterion (Martin \& Kennicutt 2001).
However, we caution that the outer disk measurements of $\Sigma_{gas}$
are lower limits in two important ways.  Most obviously, the estimate
of surface density at positions beyond $\sim R_{HII}$ is based only on
\ion{H}{1} observations, neglecting the contribution from molecular
gas traced by CO (unmapped outside of $14\arcmin\times14\arcmin$).
Also, the resolution of the \ion{H}{1} aperture synthesis map is
coarse in comparison to the scale of our detections (50$\arcsec$
vs. 5$\arcsec$).  Higher-resolution 21cm observations could plausibly
boost the apparent gas surface density in the local environment of the
UV-bright complexes.

\subsection{Underlying low surface brightness disk}

The discrete outer disk stellar complexes lie projected against a
fainter, slowly-varying field population.  To place the UV-bright
complexes in the context of M83's overall star formation history, we
have plotted the median FUV, NUV and H$\alpha$ surface brightness
(Fig. 5), plus FUV--NUV (Fig. 6), as a function of galactocentric
radius.  The underlying stellar population becomes substantially bluer
(in FUV--NUV) with increasing radius.  Dust is known to exist in the
outer disks (and even in the extended HI disks) of spiral galaxies
(Popescu \& Tuffs 2003) and thus is likely to be present in the outer
disk of M83, too, particularly given the star formation traced there
by GALEX.  Accordingly, scattered light may contribute to the emission
at each position (see also Popescu et al., this volume), tending to
smooth the UV profile.  Nevertheless, the FUV--NUV profile (Fig. 6)
shows a sequence of breaks within the bright disk of M83.  Outside of
R$_{HII}$ the FUV--NUV color becomes bluer almost monotonically,
indicating recent star formation dominates the UV emission.  The
median H$\alpha$ surface brightness profile (Fig 5.) is observed to
drop off much more rapidly than either UV band.  Given that the
dynamical timescale beyond $R_{HII}$ is $\ga 60$ Myr, the dropout of
H$\alpha$ emission compared to UV cannot likely be attributed to a
(recent) coherent decrease in the SFR at large radii.  FUV--NUV
measurements for discrete sources are overplotted on Fig. 6.  Within
$10\arcmin$ galactocentric radius, the UV-bright star-forming regions
generally appear bluer (younger) than or consistent with the FUV--NUV
color of the underlying diffuse emission.

\acknowledgments

GALEX (Galaxy Evolution Explorer) is a NASA Small Explorer,
launched in April 2003.  We gratefully acknowledge NASA's support for
construction, operation, and science analysis for the GALEX mission,
developed in cooperation with the Centre National d'Etudes Spatiales
of France and the Korean Ministry of Science and Technology.

\clearpage

\begin{figure}
\epsscale{0.9}
\plotone{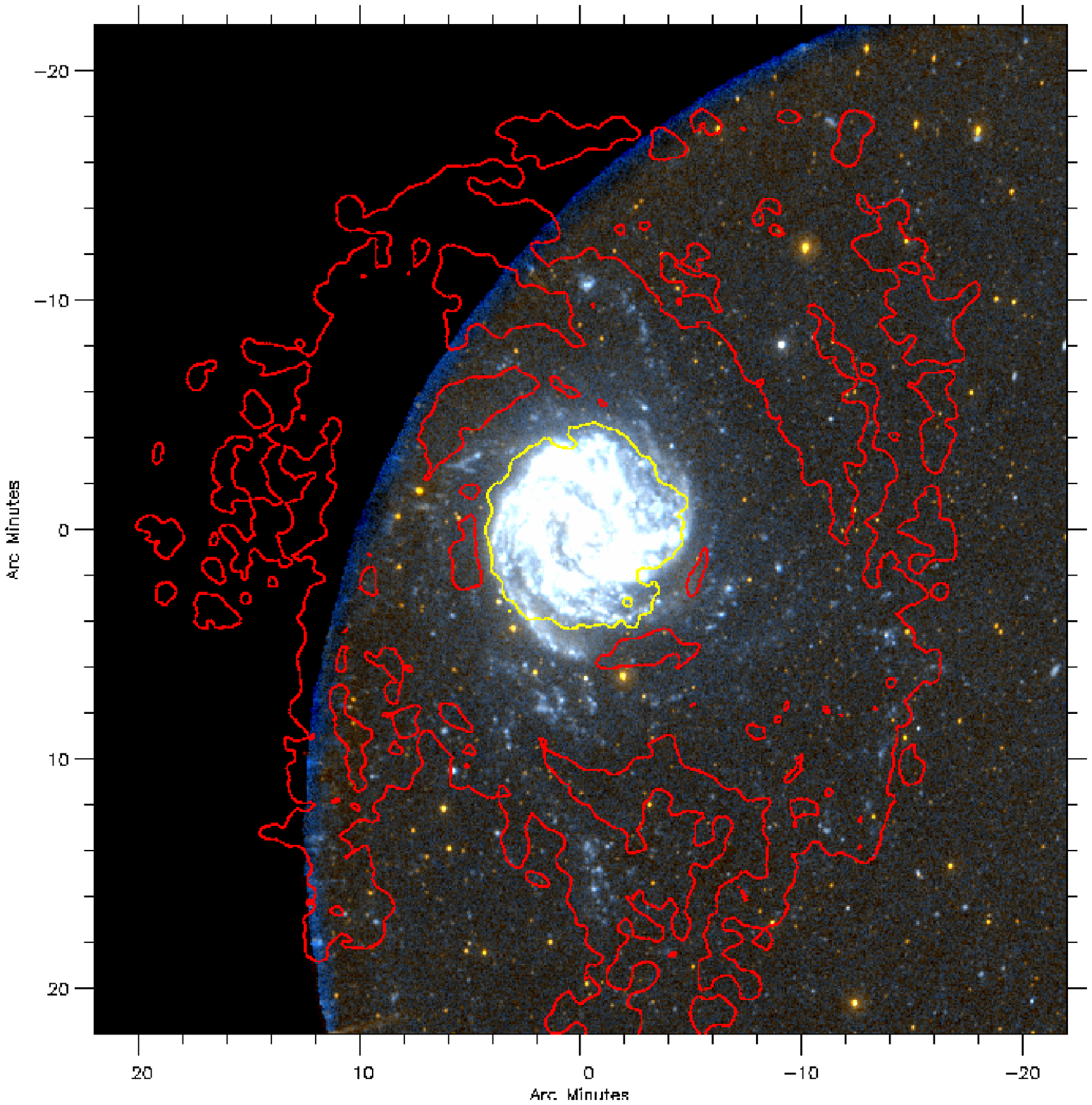}
\caption{GALEX FUV and NUV color-composite image of M83, highlighting
the newly re-discovered sites of recent star-formation in the extreme
outer disk. We show FUV in blue and NUV in red (along with their
average in green).  A red contour represents the extent of the
\ion{H}{1} distribution detected by Tilanus \& Allen (1993). The blue
\ion{H}{1} contour is drawn at $1.8\times10^{20}$ cm$^{-2}$,
corresponding to $\sim 0.4 M_\sun$ pc$^{-2}$.  We also include a
yellow contour from Crosthwaite et al. (2002), indicating total
neutral gas surface density, $\Sigma_{gas}$ of $10 M_\sun$ pc$^{-2}$.
The yellow $\Sigma_{gas}$ contour lies immediately within R$_{HII}$.
The projected spatial extent of the field ($44\arcmin$) is 57.6 kpc
(R$_{HII}$ = 6.6 kpc) at a distance of 4.5 Mpc.  }
\end{figure}

\begin{figure}
\epsscale{.85}
\plotone{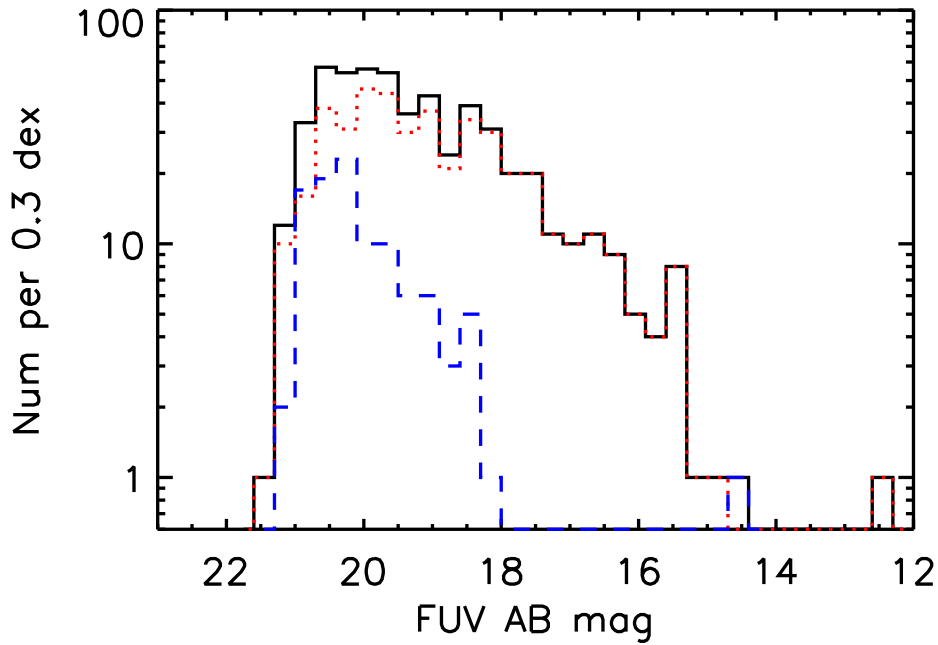}
\caption{GALEX FUV luminosity functions of the UV-bright stellar
complexes in M83, having photometric error $< 0.25$ mag.  We show the
FUV LF for all sources with a solid line, inner disk ($R < R_{HII}$)
sources with a dotted red line, and outer disk ($R \ge R_{HII}$)
sources with a dashed blue line.  Outer disk complexes appear
typically fainter and have a steeper slope in the FUV LF, relative to
inner disk objects, although detailed assessment of completeness and
blending is needed to bolster these statements.}
\end{figure}

\begin{figure}
\epsscale{.85}
\plotone{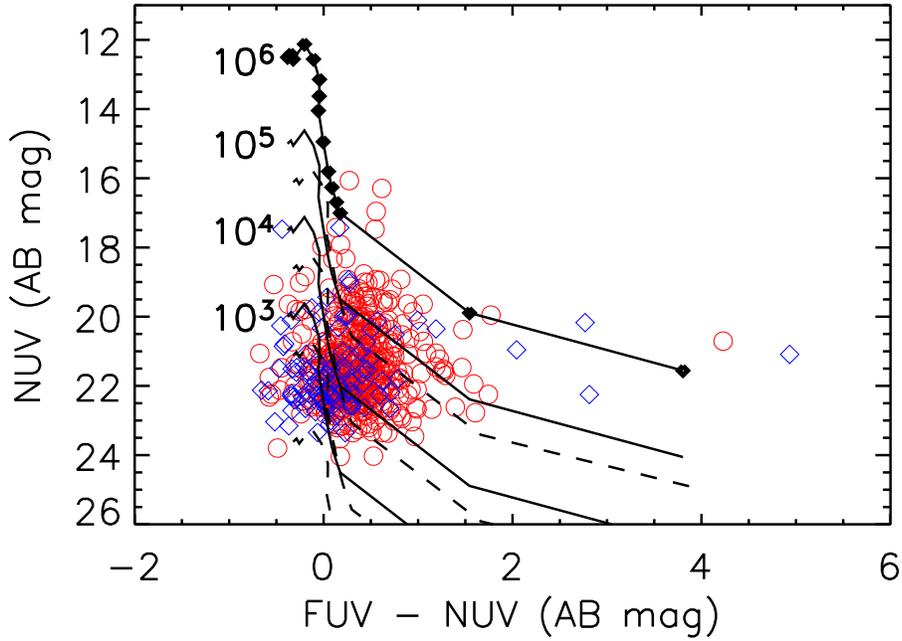}
\caption{GALEX (FUV--NUV, NUV) color-magnitude diagram for M83's
UV-bright stellar complexes (same sample and radial classification as
Fig. 2).  Inner disk sources ($R < R_{HII}$) are plotted with red
circles, whereas outer disk sources ($R \ge R_{HII}$) are indicated
with blue diamonds.  Our data points have been corrected only for
Galactic foreground extinction, E(B-V)=0.066. The black curves show
 model predictions for starburst populations of 10$^3$, 10$^4$, 10$^5$,
and 10$^6$ $M_\sun$ as a function of age (log age = 5.7, 6.0, 6.3,
6.5, 6.6, 6.75, 7.0, 7.3, 7.6, 7.75, 7.9, 8.0, 8.7, 9.0 indicated for
10$^6$ $M_\sun$).  Dashed lines indicate the same models reddened
with E(B-V)=0.5, a value slightly higher than the median E(B-V)
determined for a large sample of clusters in M83 (Harris et
al. 2001).}
\end{figure}

\begin{figure}
\epsscale{.85}
\plotone{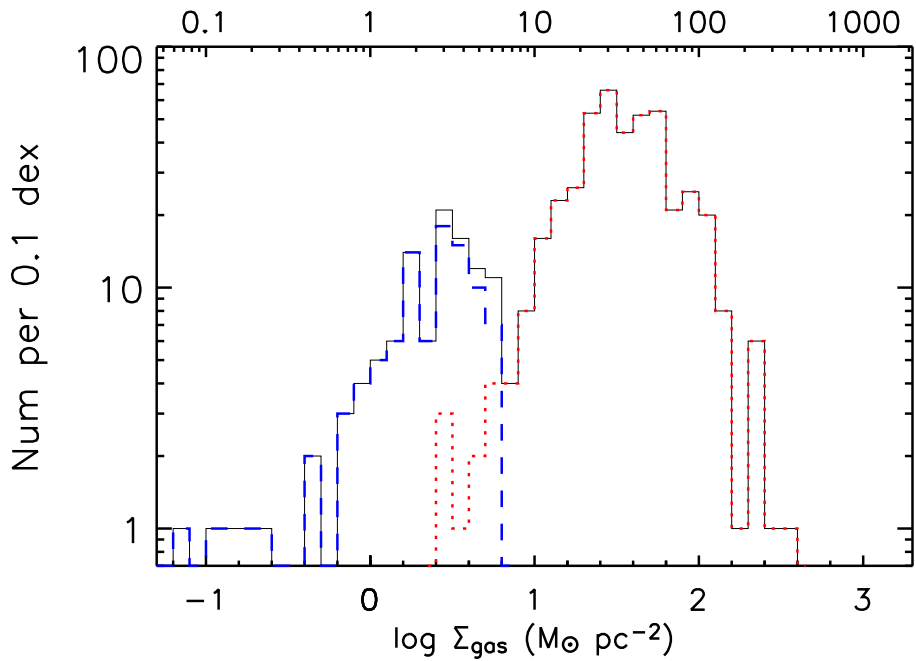}
\caption{Local $\Sigma_{gas}$ at sites of recent star formation in
M83, evaluated at 1.1 kpc (50$\arcsec$) resolution.  The overall
distribution (solid) appears largely bimodal, with inner disk UV
sources (within R$_{HII}$, dotted red) forming in locales with total
surface density greater than $\sim 10 M_\sun$~pc$^{-2}$.  Recently
formed, outer disk stellar complexes (dashed blue) exist within
environments characterized by local surface density about an order of
magnitude smaller, although the CO map used to compute the H$_2$
contribution does not extend to large galactocentric radii.
High-resolution \ion{H}{1} and CO observations toward the outer
UV-bright complexes are critically needed.  Such data would enable
investigation of \ion{H}{1} production via H$_2$ photodissociation
(Smith et al. 2000), as well as a fundamental check on the star
formation threshold (Kennicutt 1989).  We reiterate that this plot
shows local measurements of $\Sigma_{gas}$, whereas Fig. 5 presents
azimuthally averaged surface density as a function of galactocentric
radius.}
\end{figure}

\begin{figure}
\epsscale{.8}
\plotone{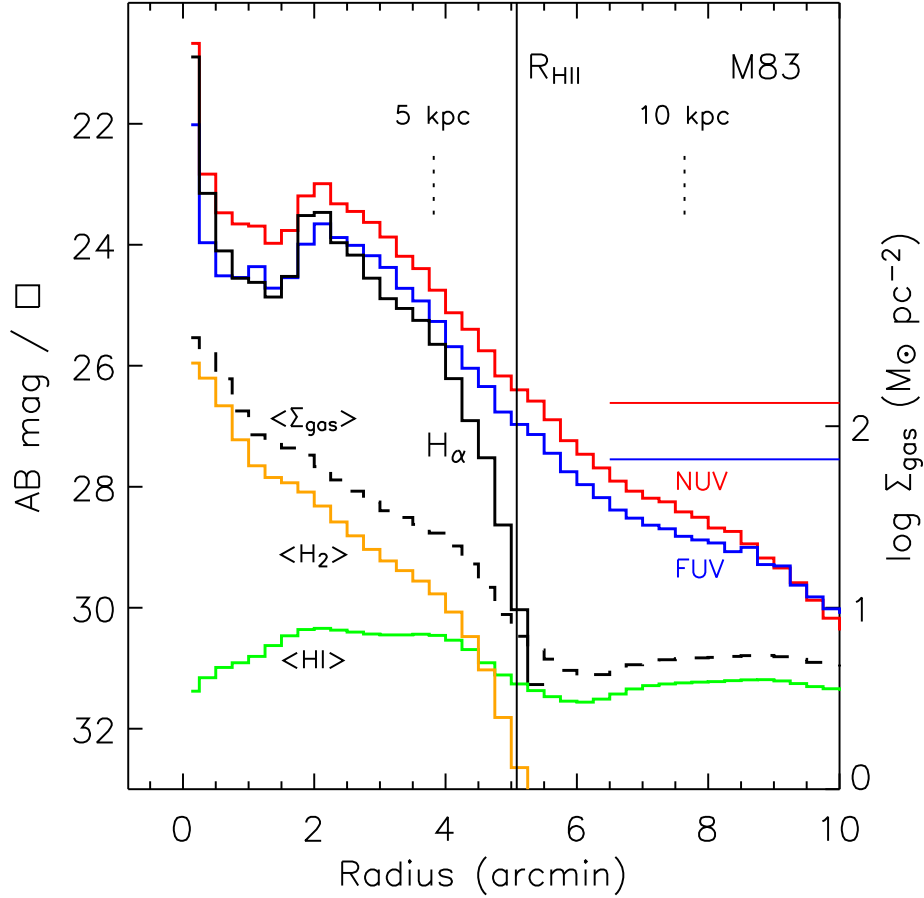}
\caption{Radial profiles of the {\it median} FUV (blue), NUV (red),
and H$\alpha$ (black) surface brightness in M83, plus {\it average}
surface densities derived from \ion{H}{1} (green) and CO (orange)
data.  Azimuthally averaged $\Sigma_{gas}$ is shown with a dashed
line.  The UV profiles are presented in absolute units, while the
H$\alpha$ profile is arbitrarily normalized.  We indicate R$_{HII}$
with a vertical line, and the UV sky background with horizontal lines.
Slight deviations from flatness across the GALEX field, coupled with
the fact that M83 is positioned on the edge of the field, make it
impractical to follow the profiles beyond $\sim 10\arcmin$. Note the
remarkable difference between the H$\alpha$ and UV profiles at
(subcritical) galactocentric radii $>5$ kpc.}
\end{figure}

\begin{figure}
\epsscale{.8}
\plotone{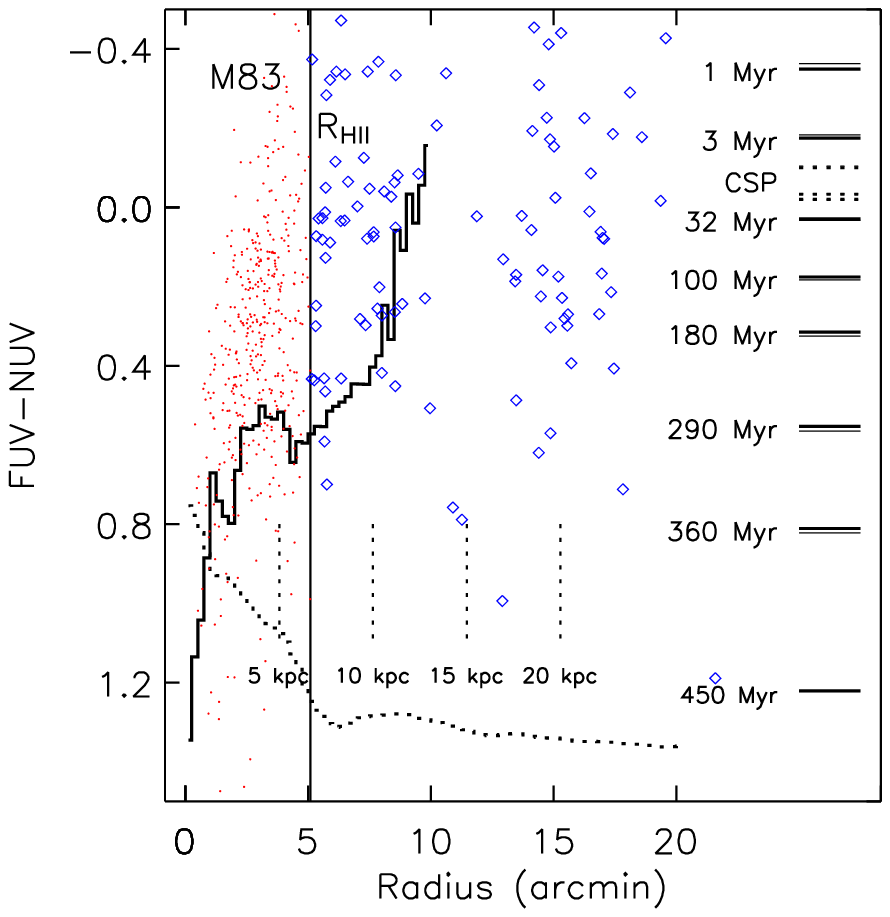}
\caption{Median FUV--NUV as a function of galactocentric radius, with
individual inner (dots) and outer (diamonds) stellar complexes plotted
for comparison.  We also show the FUV--NUV color for model starburst
populations of varied age and three different periods of continuous
star-formation (CSP: 100 Myr, 1 Gyr, and 10 Gyr).  Intrinsic and
reddened, E(B-V) = 0.2, colors are indicated with thick and thin line
segments, respectively, although the difference is minimal compared to
the observed run of FUV--NUV.  The $<\Sigma_{gas}>$ profile (dotted)
is duplicated from Fig. 5.}
\end{figure}


\begin{references}
\reference{}Allen, R. J. 1986, Nature 319, 296
\reference{}Bertin, E. \& Arnouts, S. 1996, \aaps ~117, 393
\reference{NGS} Bianchi, L. et al. 2004, in "The Local Group as an
Astrophysical Laboratory", in press (STScI, M. Livio editor) 
\reference{}Boissier, S. et al. 2003, \mnras ~321, 733
\reference{BC03}Bruzual, G. \& Charlot, G. 2003, \mnras ~344, 1000
\reference{}Crosthwaite, L. P. et al. 2002, \aj ~123, 1892
\reference{}Cuillandre, J-C. et al. 2001, \apj ~554, 190
\reference{}de Blok \& Walter 2003, \mnras ~341, 39
\reference{}Ferguson, A. M. N. et al. 1998a, \apj ~506, 19
\reference{}Ferguson, A. M. N. et al. 1998b, \aj ~116, 673
al. 1998), 
\reference{}Harris et al. 2001, \aj ~122, 3046
\reference{}Huchtmeier \& Bohnenstengel 1981, \aap ~100, 72
\reference{}Kennicutt, R. C. 1989, \apj ~344, 685
\reference{}Lelievre \& Roy 2000, \aj ~120, 1306
\reference{}Martin, C. L. \& Kennicutt, R. C. 2001, \apj ~555, 301
\reference{}Popescu, C. \& Tuffs, R. 2003, \aap ~410, 21
\reference{}Rogstad, D. H., Lockart, I. A. \& Wright, M. C. H. 1974, \apj ~193, 309
\reference{}Sellwood, J. A. \& Balbus, S. A. 1999, \apj ~511, 660 
\reference{}Smith, D. A. et al. 2000, \apj ~538, 608
\reference{}Tilanus, R. P. J. \& Allen, R. J. 1993, \aap ~274, 707
\reference{}Thim, F. et al. 2003, \apj ~590, 256
\reference{}Wamsteker, W., Lorre, J. J., and Schuster, H. E. 1983, in ``Internal Kinematics and Dynamics of Galaxies'' (Dordrecht, D. Reidel)

\end{references}
\end{document}